\documentclass[conference]{IEEEtran}

\usepackage{cite}

\usepackage{amsmath}
\usepackage{amssymb}
\usepackage{multirow}

\usepackage{array}
\usepackage{graphicx}
\usepackage{bm}
\usepackage{xcolor}
\usepackage{comment}

\definecolor{ALUColor1}{rgb}{0,0.4470,0.7410}
\definecolor{ALUColor2}{rgb}{0.8500,0.3250,0.0980}
\definecolor{ALUColor3}{rgb}{0.9290,0.6940,0.1250}
\definecolor{ALUColor4}{rgb}{0.4940,0.1840,0.5560}
\definecolor{ALUColor5}{rgb}{0.4660,0.6740,0.1880}
\definecolor{ALUColor6}{rgb}{0.3010,0.7450,0.9330}
\definecolor{ALUColor7}{rgb}{0.6350,0.0780,0.1840}
\definecolor{ALUColor8}{rgb}{0,0.7882,1}

\newif\iffull
\fulltrue

\newcommand{\eps}{\varepsilon}
\newcommand{\epsbp}{\eps_{\rm\scriptscriptstyle  BP}}
\newcommand{\epsbpuncoupled}{\eps_{\rm\scriptscriptstyle  BP,uncoupl.}}
\newcommand{\epsmapuncoupled}{\eps_{\rm\scriptscriptstyle  MAP,uncoupl.}}
\newcommand\blfootnote[1]{%
  \begingroup
  \renewcommand\thefootnote{}\footnote{#1}%
  \addtocounter{footnote}{-1}%
  \endgroup
}

\begin{document}
\title{Non-Uniformly Coupled LDPC Codes: Better Thresholds, Smaller Rate-loss, and Less Complexity}

\author{\IEEEauthorblockN{Laurent Schmalen, 
Vahid Aref, and
Fanny Jardel
}
\IEEEauthorblockA{
Nokia Bell Labs, Stuttgart, Germany. (e-mail: \{\texttt{firstname.lastname}\}\texttt{@nokia-bell-labs.com})}}
\maketitle

\begin{abstract}
We consider spatially coupled low-density parity-check codes with finite smoothing parameters. A finite smoothing parameter is important for designing practical codes that are decoded using low-complexity windowed decoders. 
By optimizing the amount of coupling between spatial positions, we show that we can construct codes with excellent thresholds and small rate loss, even with the lowest possible smoothing parameter and large variable node degrees, which are required for low error floors. 
We also establish that the decoding convergence speed is faster with non-uniformly coupled codes, which we verify by density evolution of windowed decoding with a finite number of  iterations. We also show that by only slightly increasing the smoothing parameter,  practical codes with potentially low error floors and thresholds close to capacity can be constructed. Finally, we give some indications on protograph designs.
\end{abstract}

\section{Introduction}
\label{sec:intro}
\blfootnote{The work of L. Schmalen was supported by the German Government in the frame of the CELTIC+/BMBF project SENDATE-TANDEM.}
Low-density parity-check (LDPC) codes are widely used due to their outstanding performance under low-complexity belief propagation (BP) decoding.  
However, an error probability exceeding that of maximum-a-posteriori (MAP) decoding has to be tolerated with (sub-optimal) low-complexity BP decoding. 
A few years ago, it has been empirically observed that the BP performance of 
some protograph-based, spatially coupled (SC) LDPC ensembles (also termed \emph{convolutional} LDPC codes) can improve towards the MAP performance of the underlying LDPC ensemble~\cite{Lentmaier-ita09}. Around the same time, 
this \emph{threshold saturation} phenomenon has been proven rigorously in \cite{Kudekar-it11,Kudekar-it13} for a newly introduced, \emph{randomly coupled} SC-LDPC ensemble.
 In particular, the BP threshold of that SC-LDPC ensemble tends towards its MAP threshold on any binary memoryless symmetric channel (BMS). 

SC-LDPC ensembles are characterized by two parameters: the replication factor $L$, which denotes the number of copies of LDPC codes to be places along a spatial dimension, and the smoothing parameter $w$. This latter parameter indicates that each edge of the graph is allowed to connect to $w$ neighboring spatial positions (for details, see~\cite{Kudekar-it11} and Sec.~\ref{sec:SC-LDPC}).
The proof of threshold saturation was given in the context of uniform spatial coupling and requires both $L\to\infty$ and $w\to\infty$. This poses a serious disadvantage for realizing practical codes, as relatively large structures are required to build efficient codes.

In practice, the main challenges for implementing SC-LDPC codes are the rate-loss due to termination and the decoding complexity. The rate-loss, which scales with $w$, can be made arbitrarily small by increasing $L$, however, a large $L$ can worsen the finite-length performance of SC-LDPC codes~\cite{Olmos-it15}. Known approaches to mitigate the rate-loss (e.g.,~\cite{kudekarISTC,sanatkar2016increasing}) often introduce extra structure at the boundaries, which is usually undesired. Therefore, we would like to keep the rate-loss as small as possible for a fixed, but small $L$. Additionally, the decoding complexity can be managed by employing windowed decoding (WD)~\cite{Iyengar-wd}, however, the window length and complexity scale with the smoothing parameter $w$. For both reasons, $w$ should be as small as possible, ideally $w\in\{2,3\}$, to keep the rate-loss and complexity small, e.g., in high-speed optical communications~\cite{schmalen2015spatially}.

In this paper, we construct code ensembles that have excellent thresholds for small $w$, that have smaller rate-loss than SC-LDPC ensembles and can be decoded with less complexity by maximizing the speed of the decoding wave. We achieve these properties by generalizing the uniformly coupled SC-LDPC codes of~\cite{Kudekar-it11} to allow for \emph{non-uniform} coupling. It was already recognized in~\cite{SchmalenSCC13,JardelNU} that non-uniform protographs can lead to improved thresholds in some circumstances by sacrificing a one-sided converge of the chain, which is not problematic when using WD. A very particular, exponential coupling was used in~\cite{noor2015anytime} to guarantee anytime reliability.

We extend non-uniform coupling to randomly coupled SC-LDPC ensembles and protograph-based ensembles. We analyze their performance under message passing with and without windowed decoding. We show that we can achieve excellent close-to-capacity thresholds by optimizing the coupling, for small $w$ and large $d_v$, which is required for codes with low error floors. Furthermore, we introduce a new multi-type-based non-uniform coupling that further improves the thresholds without increasing $w$. We find that the rate-loss is decreased by non-uniform coupling as well. We finally show that the decoding speed, which is an indicator of the complexity, can be increased by non-uniform coupling.

\section{Spatially Coupled LDPC Codes}
\label{sec:SC-LDPC}
We briefly describe two construction types of 
non-uniformly coupled LDPC codes: the \textit{random ensemble} and the \textit{protograph-based ensemble}. The former is easier to analyze and exhibits
the general advantages of non-uniform coupling 
while the latter is more of practical interest.

\subsection{The Random $(d_v,d_c,\protect\bm{\nu},L,M)$ Ensemble}
\label{sec:randomSC-LDPC}
We now briefly review how to sample a code from a random, non-uniformly coupled ($d_v,d_c,\bm{\nu},L,M$) SC-LDPC ensemble with regular degree distributions.
We first lay out a set of positions indexed from $z=1$ to $L$ on a \emph{spatial dimension}. 
At each spatial position (SP) $z$, there are $M$ variable nodes (VNs) and $M\frac{d_v}{d_c}$ check nodes (CNs), 
where $M\frac{d_v}{d_c} \in \mathbb{N}$ and $d_v$ and $d_c$ denote the variable and check node degrees, respectively.
The non-uniformly coupled structure is based on the smoothing distribution
$\bm{\nu}=[\nu_0,\dots,\nu_{w-1}]$ where $\nu_i>0$, 
$\sum_i\nu_i=1$ and $w>1$ denotes the smoothing (coupling) width.
The special case of $\nu_i=\frac{1}{w}$ leads to the usual, well-known spatial coupling with the uniform smoothing distribution~\cite{Kudekar-it13}.

For termination, we additionally consider $w-1$ sets of $M\frac{d_v}{d_c}$ CNs in SPs $L+1,\dots,L+w-1$. Every CN is assigned with $d_c$ ``sockets'' and imposes an even parity constraint on its $d_c$ neighboring VNs. 
Each VN in SP $z$ is connected to $d_v$ CNs in SPs $z,\ldots,z+w-1$ as follows: 
For each of the $d_v$ edges of this VN, an SP $z^\prime\in\{z,\ldots,z+w-1\}$ is randomly selected according to
the distribution $\bm{\nu}$, and then, the edge is uniformly connected to any free socket of the $Md_v$ sockets arising from the CNs in that SP $z^\prime$.
This graph represents the code with $n=LM$ code bits, distributed over $L$ SPs. Because of additional CNs in SPs $L+1,\ldots,L+w-1$, but also because of potentially unconnected CNs in SPs $1,\ldots,w-1$, the design rate is slightly decreased to $r = 1-\frac{d_v}{d_c}-\frac{1}{L}\Delta$ where
\begin{equation*}
\Delta=\frac{d_v}{d_c}
\left(w\!-\!1\!-\!\sum_{k=0}^{w-2}\left[ \left(\sum_{i=0}^k\nu_i\right)^{d_c} \!\!\!+ \left(\sum_{i=k+1}^{w-1}\nu_i\right)^{d_c}\right]\right)
\end{equation*}
which increases linearly with $w$. 

In the limit of $M$, the asymptotic performance of this ensemble on a binary erasure channel (BEC) can be analyzed using density evolution, with
\begin{equation}
\label{eq:de}
x_{z}^{(t+1)} = \eps\!\left(\!1\!-\!\sum_{i=0}^{w-1}\nu_i\left(1\!-\!\sum_{j=0}^{w-1}\nu_j x_{z+i-j}^{(t)}\right)^{d_c-1}\right)^{d_v-1}
\end{equation}
where $\eps$ denotes the channel erasure probability and $x_z^{(t)}$ the average erasure probability of the outgoing messages from VNs in SP $z$ at iteration $t$. The messages are initialized as $x_z^{(0)} = \eps$, if $z\in[1,L]$ and $x_z^{(0)} = 0$ otherwise. For $\nu_i=\frac{1}{w}$, \eqref{eq:de} becomes the known 
DE equation for SC-LDPC codes with uniform coupling~\cite[Eq.~(7)]{Kudekar-it11}.

\subsection{Protograph-based SC-LDPC Ensembles}
\label{sec:protoSC-LDPC}

\begin{figure}[tb!]
\centering
\includegraphics{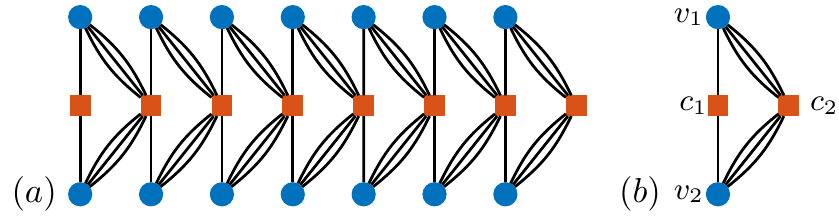}
\caption{\label{fig:proto} Protograph of a $(d_v=4,d_c=8,L=7,w=2)$ SC-LDPC ensemble with non-uniform coupling $(b)$ The elementary segment of the chain denoted by the 3-tuple $(4,1,1)$.}
\end{figure}

SC-LDPC ensembles with a certain predefined structure
can be constructed by means of \textit{protographs}~\cite{mitchell2015spatially}. The Tanner graph of the protograph-based SC-LDPC code is some $M$-cover of the protograph, i.e.,
$M$ copies of the protograph are bound together by random permutation of the edges between the same type of sockets. Protograph-based SC-LDPC codes are of practical interest because of their simple hardware implementation and their excellent 
finite-length performance~\cite{stinner2016waterfall}.
An exemplary protograph of an SC-LDPC code with non-uniform coupling is shown in Fig.~\ref{fig:proto}-a). As the coupled protograph is a chain of repeating segments, we represent coupled protographs by their distinct elementary segment shown in Fig.~\ref{fig:proto}-b). We use the 3-tuple $(d_v,b_1,b_2)$ to describe the elementary segment, with $d_v$ the regular variable node degree, $b_1$ the number of parallel edges between VN $v_1$ and CN $c_1$ and $b_2$ the number of parallel edges between VN $v_2$ and CN $c_1$. 

\subsection{Windowed Decoder Complexity}
\label{sec:speedWD}
The decoding complexity is an important parameter for practical SC-LDPC codes.
Consider the profile of densities $[x_0^{(t)},x_1^{(t)},\dots]$ in \eqref{eq:de}. It has been shown in~\cite{Kudekar-it11,kudekar2015wave} that the profile behaves like a ``wave'': it shifts along the spatial dimension with ``a constant speed'' as the BP decoder iterates. The wave propagation speed is analytically analyzed and bounded in~\cite{aref2013convergence},\cite{el2016velocity}.

The wave-like behaviour enables efficient \textit{sliding windowed decoding}~\cite{Iyengar-wd}: the decoder updates the BP messages of edges lying in a window of $W_D$ SPs $I$ times, and then shifts the window one SP forward and repeats. Thus, the decoding complexity scales with $O(W_DILMd_v)$ as there are $2MLd_v$ BP messages and each BP message is updated $W_DI$ times.

The required window size $W_D$ is an increasing function of the smoothing factor $w$~\cite{Iyengar-wd} which implies that we should keep $w$ small. The number of iterations $I>\frac{1}{v}$ where $v$ is the speed of the wave. In the continuum limit of the spatial dimension, $v$ is defined as the amount displacement of the profile along the spatial dimension after one iteration. For the discrete case of \eqref{eq:de}, the speed can be estimated by 
\begin{equation}
\label{eq:VD}
v\approx v_D=\frac{D}{T_D},
\end{equation}
where $T_D$ in the minimum number of iterations required for the displacement of the profile by more than $D$ SPs, i.e.,
\begin{equation*}
T_D = \min\{T\in\mathbb{N}\mid x_z^{(t+T)}\leq x_{z-D}^{(t)}, \;{\rm for}\;t>0\;\wedge\;z\leq \lfloor L/2\rfloor \}.
\end{equation*}
The approximation of $v$ becomes more precise by choosing larger $D$. We chose $D=10$ in this paper. 

We quickly recapitulate the asymptotic analysis for the windowed decoder here. Instead of the windowed decoder proposed in~\cite[Def. 4]{Iyengar-wd}, we employ a slightly modified, more practical version, which updates the complete window after one decoding step. 
For every windowed decoding step, indexed by $c\in[1,L]$, we generate a copy $\bm{y}_{c,z}^{(0)}$ of the vector $\bm{x}=(x_1^{(c-1)},\ldots,x_{L+w-1}^{(c-1)})$ on which we apply the update rule~\eqref{eq:de} for SPs $z\in\{c,c+1,\ldots,c+W_D-1\}$ only, for a total of $I$ iterations. After $I$ iterations, we update the SPs as
\[
x_z^{(c)} = \left\{\begin{array}{ll}
x_z^{(c-1)} & \text{if }z \not\in[c,c+W_D) \\
y_{z-c+1}^{(I)} &\text{otherwise}\end{array}\right.
\]
We use a finite number of iterations in the windowed decoder to accurately predict the performance of a practical decoder.

\section{Non-Uniform Coupling: Random Ensembles}
\label{sec:randomEns}
In this section, we optimize non-uniformly SC-LDPC ensembles with random coupling for the BEC. First, we consider $w=2$, the smallest possible smoothing parameter. This case has a high practical interest as $w$ should be kept as small as possible in order to keep the decoding latency and window length $W_D$ manageable when employing windowed decoding. We show numerically that non-uniform coupling improves the BP threshold  and also the decoding complexity as the total number of iterations decreases. Afterwards, we show the advantages of non-uniform coupling $w>2$.

\subsection{Non-Uniform Unit-Memory Coupling ($w=2$)}
\begin{figure}[b!]
\includegraphics{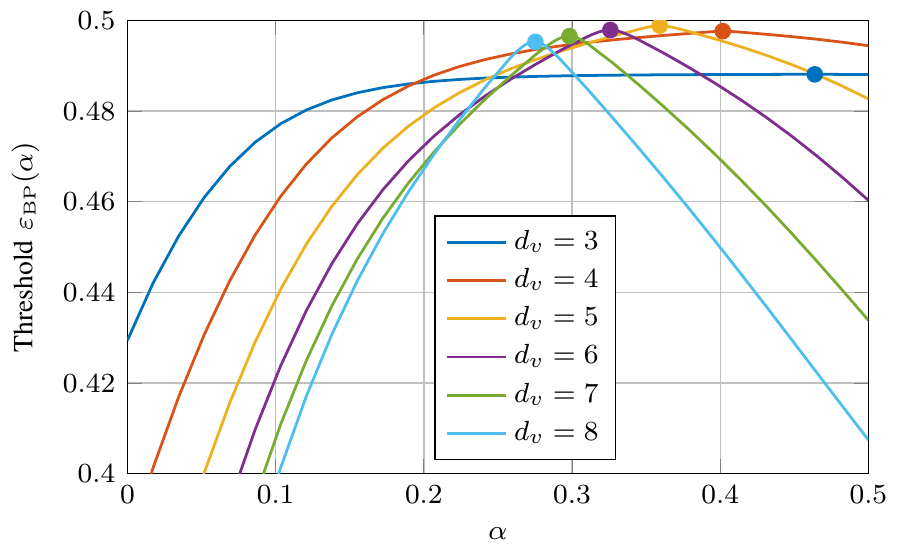}
\caption{Thresholds for the $(d_v,2d_v,[\alpha,1-\alpha],L=100)$ ensemble (with $w=2$ and rate $\approx \frac{1}{2}(1-\frac{1}{L})$).}
\label{fig:reg_thresholds}
\end{figure}

Consider a random $(d_v,d_c,\bm{\nu},L,M)$ SC-LDPC ensemble
with smoothing vector $\bm{\nu}=[\alpha,1-\alpha]$. It is enough to  assume $0\leq\alpha\leq\frac{1}{2}$ because of symmetry. In the limit of $M$, the  asymptotic performance of the ensemble over BEC can be evaluated using DE. We consider the BP threshold 
\[
\epsbp(\alpha) = \sup\{\eps : x_z^{(\ell)} \to 0\text{ as }\ell\to\infty,\forall z\in[1,L]\},
\]
where $x_z^{(\ell)}$ is updated according to \eqref{eq:de}. 
Figure~\ref{fig:reg_thresholds} illustrates
$\epsbp(\alpha)$ in terms of $\alpha$
for different values of $d_v$.
Each curve has two minima and a maximum. 
The two minima are at $\alpha=0$ and $\alpha=\frac{1}{2}$ where
$\epsbp(\alpha=0)=\epsbpuncoupled$ corresponds to
the BP threshold of the uncoupled ensemble and $\epsbp(\alpha=\frac{1}{2})$ corresponds to the BP threshold of the SC-LDPC ensemble with uniform coupling.
The respective maxima of the curves are indicated by a marker and obtained for $\alpha^*$.
We can see that uniform coupling ($\alpha=1/2$) does not lead to the best thresholds. In particular, if we increase $d_v$, which is required for constructing codes with very low error floors, uniform coupling with $w=2$ is not efficient anymore, and the thresholds are significantly away from the BEC capacity. With an optimized $\alpha^\star$, we can achieve thresholds that are close to capacity (and the MAP threshold of the uncoupled LDPC ensemble $\epsmapuncoupled$) and significantly outperform the uncoupled and the uniformly coupled cases. Table~\ref{tab:reg_thresholds} gives the thresholds of the optimized codes together with the unoptimized, uniformly coupled and uncoupled cases. Although coupling always improves the threshold, with $w=2$, uniform coupling is not a good solution and significantly better thresholds are obtained by non-uniform coupling, especially for larger $d_v$. Moreover, it is easy to show that the rate-loss $\Delta$ is maximized for uniform coupling ($\alpha=1/2$). Hence non-uniform coupling will always reduce the rate-loss.
We can see that as $d_v$ increases, $\alpha^\star$ decreases as well. An interesting open question is whether $\alpha$ saturates to some constant or if it will converge to zero.

\begin{table}
\centering
\caption{Optimal $\alpha^\star$ and the BP and MAP thresholds of the uncoupled codes, $\epsbpuncoupled$ and $\epsmapuncoupled$, and BP thresholds of the uniformly coupled codes, $\epsbp(\alpha=1/2)$, and with optimized $\alpha^\star$ for different rate $\approx \frac{1}{2}(1-\frac{1}{L})$ with regular VN degree $d_v$.}\label{tab:reg_thresholds}
\begin{tabular}{@{}cc@{\;\;\;\;}c@{\;\;\;\;}c@{\;\;\;\;}cc@{}}
$d_v$ & $\alpha^\star$ & $\epsbpuncoupled$ & $\epsmapuncoupled$ & $\epsbp(\alpha=1/2)$ & $\epsbp(\alpha^\star)$ \\
\hline
\\[-0.9ex]
3 & 0.4517 & 0.4294 & 0.48815 & 0.4880(8) & 0.4881(0)\\
4 & 0.4017 & 0.3834 & 0.49774 & 0.4944 & 0.4976\\
5 & 0.3590 & 0.3415 & 0.49949 & 0.4827 & 0.4989\\
6 & 0.3252 & 0.3075 & 0.49988 & 0.4603 & 0.4979\\
7 & 0.2978 & 0.2798 &0.49997 & 0.4338 & 0.4965\\
8 & 0.2745 & 0.2570 & 0.49999 & 0.4074 & 0.4953\\
9 & 0.2544 & 0.2378 & 0.49999 & 0.3829 & 0.4943\\
10 & 0.2368 & 0.2215 & 0.49999 & 0.3606 & 0.4936\\
\end{tabular}
\end{table}

\begin{figure}[b!]
\includegraphics{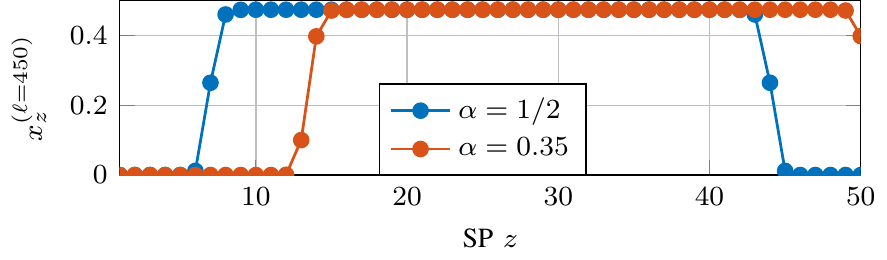}
\caption{Average message erasure probability $x_z^{(\ell)}$ for the $(5,10,\alpha,50)_{w=2}$ ensemble for $\alpha\in\{0.35,1/2\}$ and with $\ell=450$ for $\epsilon=0.48$.}
\label{fig:wave_example}
\end{figure}

\begin{figure}[tb!]
\includegraphics{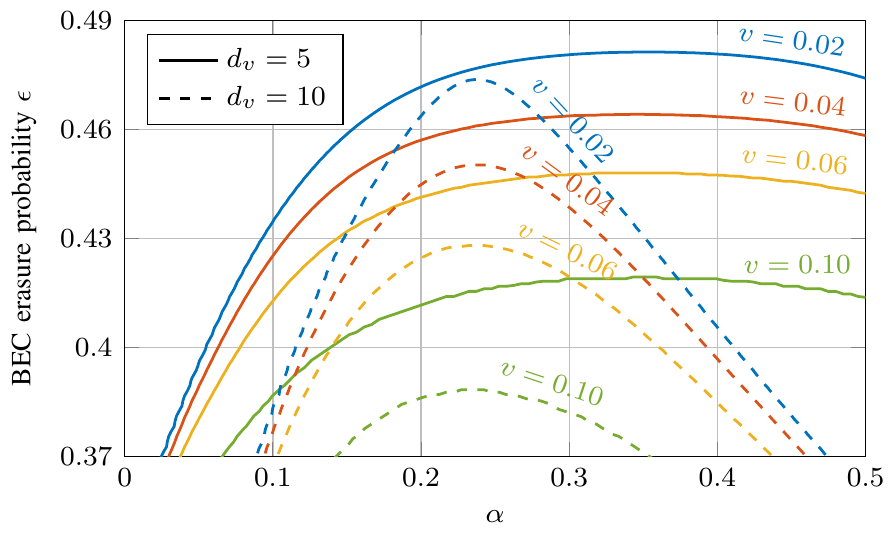}
\caption{\label{fig:reg_speed5_and10} Decoding speed contour lines for the random $(5,10,[\alpha,1-\alpha],L=100)$ 
SC-LDPC ensemble (solid curves) and 
the $(10,20,[\alpha,1-\alpha],L=100)$
SC-LDPC ensemble (dashed curves) 
with design rate $\approx\frac{1}{2}(1-\frac{1}{L})$.
}
\end{figure}

Non-uniform coupling can also decrease the decoding complexity of windowed decoding.
Figure~\ref{fig:wave_example} illustrates the effect of non-uniform coupling on the wave propagation. While uniform coupling ($\alpha=\frac{1}{2}$) leads to a wave propagation from both ends towards the middle, non-uniform coupling sacrifices one of those waves in favor of the other one, which will (usually) travel at a faster velocity.
We compute the speed $v$ according to~\eqref{eq:VD} for different values of $\alpha\in[0,1/2]$ and different values of $\eps\in[\epsbp(\alpha=0),\epsbp(\alpha^\star)]$ and show the contour lines of equal decoding speed $v$ in Fig.~\ref{fig:reg_speed5_and10} 
for $d_v=5$ and $d_v=10$. Points along a contour line indicate that the decoding wave moves with the same speed. When building practical decoders, usually a hardware constraint is imposed which limits the amount of operations that can be done. Hence also the decoding speed is limited. We can see that for a fixed speed $v$, non-uniformly coupled codes can be operated at much higher erasure probability than with uniform coupling. Note that the maxima of the speed contours coincide practically with the $\alpha^{\star}$ maximizing the threshold.

\begin{figure}[tb!]
\includegraphics{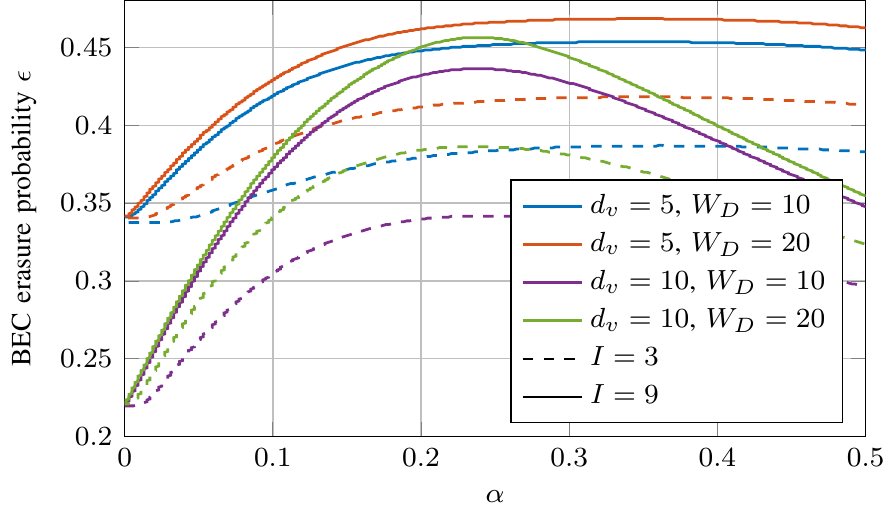}
\caption{Windowed decoding thresholds for the $(5,10,[\alpha,1-\alpha],L=100)$ and $(10,20,[\alpha,1-\alpha],L=100)$ ensembles for different window configurations as a function of $\alpha$.}
\label{fig:reg_windowed10}
\end{figure}

Figure~\ref{fig:reg_speed5_and10} suggests that windowed decoding also benefits from non-uniform coupling. For this reason, we use density evolution including windowed decoding, as detailed in Sec.~\ref{sec:speedWD}. Figure~\ref{fig:reg_windowed10} exemplarily  shows the thresholds for windowed decoding for the $(5,10,[\alpha,1-\alpha],L=100)$ and the $(10,20,[\alpha,1-\alpha],L=100)$ SC-LDPC ensembles for four window configurations: $W_D\in\{10,20\}$ and $I\in\{3,9\}$. We see a good agreement between the speed contour lines of Fig.~\ref{fig:reg_speed5_and10} and the windowed decoding thresholds. Again we can see that for non-uniformly coupled codes and identical window configurations, we can significantly increase the decoding threshold.

\subsection{Non-Uniform Coupling with $w > 2$}

We have seen in the previous section that
non-uniform coupling can increase the BP threshold
if we constrain $w=2$. However, for $d_v>5$, we have to tolerate a gap to capacity.
In this case, we can relax the constraint on $w$. In fact, for $w>2$, non-uniform coupling can be more beneficial as there are more degrees of freedom for optimizing the smoothing vector $\bm{\nu}$. We numerically show in the following that it results in a faster saturation of the BP threshold to capacity even for small values of $w$, e.g., $w=3$. 

Consider the DE equation \eqref{eq:de} for a random $(d_v,d_c,\bm{\nu},L)$ SC-LDPC ensemble over a BEC. Let $\bm{\nu}=[\nu_1,\nu_2,1-\nu_1-\nu_2]$ with $w=\text{dim}(\bm{\nu})=3$.
For regular ensembles with asymptotic rate $r=\frac{1}{2}$ ($d_c=2d_v$),
we observe that the BP threshold, $\epsbp(\bm{\nu})$, depends on
the choice of $\bm{\nu}$ and can get very close to the capacity. 
We used a grid search with a fine resolution to 
numerically optimize the BP threshold for the ensembles with $d_v\in\{4,\ldots, 10\}$. The results are given in Tab.~\ref{tab:reg3_thresholds} where 
the optimized smoothing distribution is denoted by $\bm{\nu}^\star=[\nu_1^\star,\nu_2^\star,1-\nu_1^\star-\nu_2^\star]$. We observe that the BP thresholds almost saturate to the capacity (or $\epsmapuncoupled$, respectively), while the BP threshold of uniformly coupled ensembles ($\epsbp(\bm{\nu}=[\frac{1}{3},\frac{1}{3},\frac{1}{3}])$) have a gap to capacity which increases for larger $d_v$. Note that especially for small $d_v$, many different choices of $\bm{\nu}$ lead to good thresholds $\epsbp$. In that case, we select the optimum $\bm{\nu}^\star$ which leads to a good threshold and also yields a small rate loss $\Delta$. Note that in contrast to the $w=2$ case, where the rate-loss was maximal for uniform coupling,  it is not hard to show that the rate-loss $\Delta$ for $w=3$ is maximized with $\bm{\nu} = [\frac{1}{2},0,\frac{1}{2}]$. It is an interesting open question whether it is possible to construct capacity-achieving codes with a finite $w$. 

\begin{table}[t!]
\centering
\caption{Optimal $\bm{\nu}$ and the BP thresholds of the uniformly coupled codes $\epsbp(\bm{\nu}=[\frac{1}{3},\frac{1}{3},\frac{1}{3}])$ and with optimized $\bm{\nu}^\star=[\nu_1^\star,\nu_2^\star,1-\nu_1^\star-\nu_2^\star]$  for the random $(d_v,2d_v,\bm{\nu},L=100)$ SC-LDPC ensembles with rate $\approx \frac{1}{2}(1-\frac{2}{L})$, as well as rate-losses $\Delta$ for both uniform and non-uniform coupling.}\label{tab:reg3_thresholds}
\begin{tabular}{@{}*{6}{c@{\;\;\;\,}}c@{}}
$d_v$ & $\nu_1^\star$ & $\nu_2^\star$& $\epsbp([\frac{1}{3},\frac{1}{3},\frac{1}{3}])$ & $\epsbp(\bm{\nu}^\star)$ & $\Delta([\frac{1}{3},\frac{1}{3},\frac{1}{3}])$ & $\Delta(\bm{\nu}^\star)$ \\[0.4ex]
\hline
\\[-1.4ex]
3 & 0.0789 & 0.4737 &  0.48815&  0.4881(5) & 0.911 & 0.676 \\
4 & 0.1842 & 0.4211 &  0.4977 &  0.4977(4) & 0.961 & 0.893\\
5 & 0.2632 & 0.2105 &  0.4989 &  0.4994(7) & 0.983 & 0.975\\
6 & 0.2465 & 0.1496 &  0.4967 &  0.4998(7) & 0.992 & 0.982\\
7 & 0.2355 & 0.1247 &  0.4904 &  0.4999(7) & 0.997 & 0.987\\
8 & 0.2244 & 0.1025 &  0.4797 &  0.4999(8) & 0.998 & 0.991\\
9 & 0.2147 & 0.0803 &  0.4652 &  0.4999(5) & 0.999 & 0.993\\
10 & 0.2063 & 0.0665&  0.4486 &  0.4999(4) & 1.000 & 0.994
\end{tabular}
\end{table}

\subsection{Non-Uniform Coupling with Different Types}
Non-uniform coupling is a general concept. So far, 
we presented the simplest way of non-uniform coupling in which the edges of all VNs in an SP are randomly connected 
according to a distribution $\bm{\nu}$. Generally, the edges of each VN 
can be connected according to a set of distributions. 
Let us illustrate the benefits of such coupling by an example. Consider again a coupled LDPC ensemble with $w=2$ and $d_c=2d_v$. Inspired by the protograph structure shown in Fig.~\ref{fig:proto}, we partition the VNs in each SP into two sets of equal size,  called ``upper set'' and ``lower set''. As described in Sec.~\ref{sec:randomSC-LDPC}, the edges of VNs in the upper set are randomly connected to CNs according to
the ``upper'' smoothing distribution $\overline{\bm{\nu}}=[\overline{\alpha},1-\overline{\alpha}]$. Similarly, the edges of VNs in the lower set are distributed according to the ``lower'' smoothing distribution $\underline{\bm{\nu}}=[\underline{\alpha},1-\underline{\alpha}]$. Therefore, each CN receives two types of BP messages from VNs. Let $\overline{x}_z^{(t)}$ ($\underline{x}_z^{(t)}$) denote the average erasure probability of the BP messages flowing from VNs of the upper set (lower set) in SP $z$ at iteration $t$. Then the DE equations become

{\small
\begin{align*}
&\overline{y}_z^{(t)}\!=\!\left(1\!-\!(\overline{\alpha}\overline{x}_z^{(t)} \!+\! (1\!-\!\overline{\alpha})\overline{x}_{z-1}^{(t)})\right)^{d_v\!-\!1}\!\!\left(1\!-\!(\underline{\alpha}\underline{x}_z^{(t)} \!+\! (1\!-\!\underline{\alpha})\underline{x}_{z-1}^{(t)})\right)^{d_v}\\
&\underline{y}_z^{(t)}\!=\!\left(1\!-\!(\overline{\alpha}\overline{x}_z^{(t)} \!+\! (1\!-\!\overline{\alpha})\overline{x}_{z-1}^{(t)})\right)^{d_v}\!\!\left(1\!-\!(\underline{\alpha}\underline{x}_z^{(t)} \!+\! (1\!-\!\underline{\alpha})\underline{x}_{z-1}^{(t)})\right)^{d_v\!-\!1}\\
&\overline{x}_z^{(t+1)}=\eps\left(1-( \overline{\alpha}\overline{y}_z^{(t)} + (1-\overline{\alpha})\overline{y}_{z+1}^{(t)})\right)^{d_v-1}\\
&\underline{x}_z^{(t+1)}=\eps\left(1-( \underline{\alpha}\underline{y}_z^{(t)} + (1-\underline{\alpha})\underline{y}_{z+1}^{(t)})\right)^{d_v-1}
\end{align*}}%
Using DE analysis and a rough exhaustive search, we optimized $\overline{\alpha}$ and $\underline{\alpha}$ to find the largest BP threshold for different values of $d_v$.
The thresholds are summarized in Tab.~\ref{tab:reg4_thresholds}. We observe that the thresholds almost saturate to capacity for $d_v=6$ and $d_v=7$ with only $w=2$.

\begin{table}
\centering
\caption{Non-uniform coupling with two types: BP thresholds of random non-uniformly coupled ensembles with $d_c=d_v$, $w=2$, $L=100$
and optimal $[\overline{\alpha}^\star,1-\overline{\alpha}^\star]$
and $[\underline{\alpha}^\star,1-\underline{\alpha}^\star]$.\label{tab:reg4_thresholds}}
\begin{tabular}{@{}c@{\quad\qquad}c@{}}
\begin{tabular}{@{}cccc@{}}
$d_v$ & $\overline{\alpha}^\star$ & $\underline{\alpha}^\star$& $\epsbp$\\
\hline
\\[-0.9ex]
5 &0.350 &0.362 &0.4989\\
6 &0.278 &0.375 &0.4998\\
7 &0.248 &0.349 &0.4998\\
\end{tabular} &
\begin{tabular}{@{}cccc@{}}
$d_v$ & $\overline{\alpha}^\star$ & $\underline{\alpha}^\star$& $\epsbp$\\
\hline
\\[-0.9ex]
8 &0.227 &0.323 &0.4996\\
9 &0.209 &0.300 &0.4995\\
10 &0.195 &0.279 &0.4994\\
\end{tabular}
\end{tabular}
\end{table}

\section{Non-Uniform Coupling: Protograph Ensembles }
\label{sec:protEns}

\begin{figure}[t!]
\centering
\includegraphics{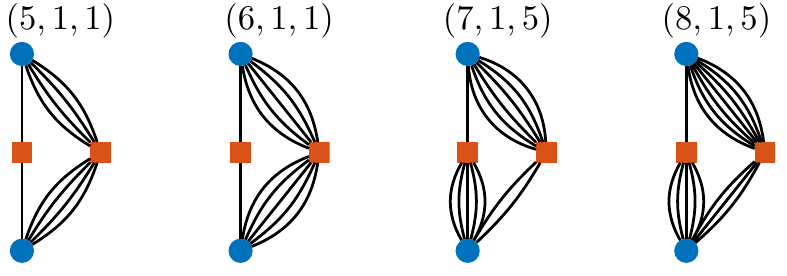}
\caption{Example of optimized protographs, represented by their elementary segment, with various unequal coupling for $d_v\in\{5,6,7,8\}$.}
\label{fig:protographs_considered}
\end{figure}

As most practical codes are based on protographs, we extend the findings of this paper to protograph-based codes with the elementary building segment of Fig.~\ref{fig:proto}-b). In comparison to the random ensembles, there is less room for optimization
as there are finite choices for $b_1$ and $b_2$, each requiring a
separate DE analysis, which is also slightly more complicated as the BP messages come from different edge types (multi-edge types DE).
We computed DE thresholds for all possible protographs based on a simple elementary segment  with 2 VNs and 2 CNs for $L=100$ ($r=0.495$). In Tab.~\ref{tab:proto_thresholds}, we summarize the best protographs and the respective thresholds that we find for different choices of $d_v$. Some of the best elementary segments are shown in Fig.~\ref{fig:protographs_considered}. Up to $d_v=6$, protographs with $b_1=b_2=1$ are optimal, however, when $d_v > 6$, interestingly, the choice $b_1=1$ and $b_2=5$ becomes optimal. 

\begin{table}[b!]
\caption{Thresholds of non-uniformly coupled regular protograph ensembles with simple elementary protographs.}\label{tab:proto_thresholds}
\centering
\begin{tabular}{cc}
\begin{tabular}{cc}
$(d_v,b_1,b_2)$ & Threshold $\eps_{\rm BP}$ \\
\hline
\\[-1.2ex]
$(3,1,1)$ & 0.48815\\
$(4,1,1)$ & 0.49741\\
$(5,1,1)$ & 0.49811\\
$(6,1,1)$ & 0.49667\\
$(7,1,5)$ & 0.49257\\
$(8,1,5)$ & 0.49451\\
$(9,1,5)$ & 0.49627\\
$(10,1,5)$ & 0.49711\\
\end{tabular} & \begin{tabular}{cc}
$(d_v,b_1,b_2)$ & Threshold $\eps_{\rm BP}$ \\
\hline
\\[-1.2ex]
$(11,1,5)$ & 0.49693\\
$(12,1,5)$ & 0.49612\\
$(13,1,5)$ & 0.49502\\
$(14,1,5)$ & 0.49377\\
$(15,1,5)$ & 0.49246\\
$(16,1,5)$ & 0.49113\\
$(17,1,5)$ & 0.48981\\
$(18,1,5)$ & 0.48850\\
\end{tabular}
\end{tabular}
\end{table}

\begin{figure}[t!]
\includegraphics{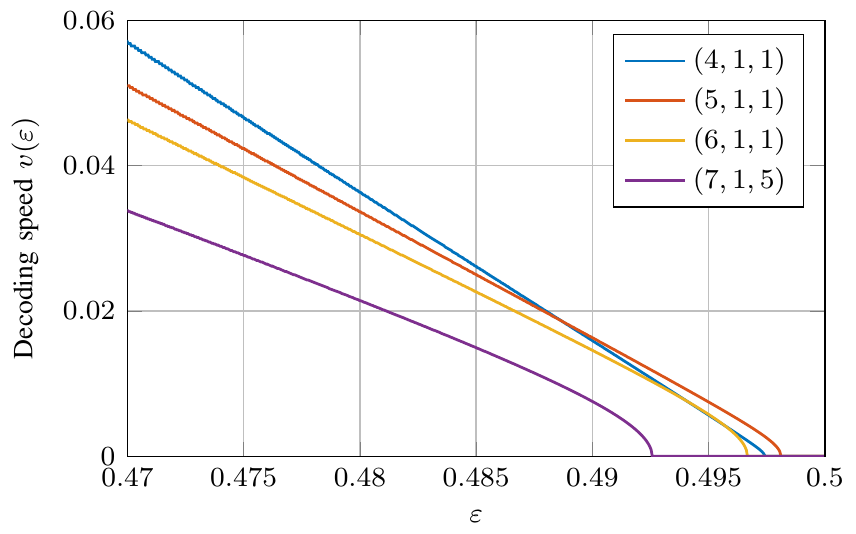}
\caption{Decoding speeds for different optimal protographs.}
\label{fig:proto_speed}
\end{figure}

In Fig.~\ref{fig:proto_speed}, we plot the decoding speeds for the best protographs with $d_v\in\{4,5,6,7\}$. We can see that for $\eps < 0.488$, the protograph $(4,1,1)$ has the highest decoding speed and thus leads to the smallest decoding complexity, while for $\eps \geq 0.488$, the protograph $(5,1,1)$ has the highest speed due to its different slope. Using an \emph{exhaustive} search over \emph{all possible} elementary protograph segments with 2 VNs and 2 CNs and with $d_v\leq 18$, we have verified that these two protographs are indeed the ones yielding the highest overall speeds and are good candidates for implementation.

\section*{Acknowledgments}
The authors would like to acknowledge R{\"u}diger Urbanke and Shrinivas Kudekar for interesting discussions and suggestions leading to the work in this paper and its presentation.


\end{document}